\documentclass[11pt]{article}
\usepackage{graphicx,cite,amsfonts}
\usepackage[left=2cm,right=2cm,bottom=2cm,includeheadfoot,a4paper]{geometry}
\usepackage[onehalfspacing]{setspace}

\def\rv{{\bf r}}
\def\bv{{\bf b}}

\def\kv{{\bf k}}

\def\(#1){(\ref{#1})}           
\newcommand{\Rset}{\mathbb{R}}
\newcommand{\Zset}{\mathbb{Z}}

\begin{document}

\title{Soft quasicrystals - Why are they stable?}  

\author{Ron Lifshitz$^\dagger$ and Haim Diamant$^*$\\ $^\dagger$School
  of Physics \& Astronomy and $^*$School of Chemistry\\ Raymond and
  Beverly Sackler Faculty of Exact Sciences\\ Tel Aviv University, Tel
  Aviv 69978, Israel}
\date{October 31, 2006}
\maketitle

\begin{abstract}
  In the last two years we have witnessed the exciting experimental
  discovery of soft matter with nontrivial quasiperiodic long-range
  order---a new form of matter termed a \emph{soft quasicrystal}. Two
  groups have independently discovered such order in soft matter: Zeng
  \emph{et al.}  [Nature {\bfseries 428} (2004) 157] in a system of
  dendrimer liquid crystals; and Takano \emph{et al.}  [J.\ Polym.\ 
  Sci.\ Polym.\ Phys.\ {\bfseries 43} (2005) 2427] in a system of ABC
  star-shaped polymers. These newly discovered soft quasicrystals not
  only provide exciting platforms for the fundamental study of both
  quasicrystals and of soft matter, but also hold the promise for new
  applications based on self-assembled nanomaterials with unique
  physical properties that take advantage of the quasiperiodicity,
  such as complete and isotropic photonic band-gap materials. Here we
  provide a concise review of the emerging field of soft
  quasicrystals, suggesting that the existence of two natural
  length-scales, along with 3-body interactions, may constitute the
  underlying source of their stability.
 \end{abstract}

\section{Background}

The discovery of quasicrystals by Shechtman more than two decades
ago~\cite{shechtman} signaled the beginning of a remarkable scientific
revolution~\cite{cahn}, in which some of the most basic notions of
condensed matter physics have undergone a thorough reexamination.
Today, the science of quasicrystals, with its growing number of
textbooks~\cite{steinhardt,janot,stadnik,suck,senechal,%
  moody,baakemoody,dubois}, is in its adolescence. Old paradigms are
being carefully transformed into new ones~\cite{rebirth}; definitions
are being changed~\cite{quasidef}; space-group theory has been
generalized to quasicrystals using two alternative
approaches~\cite{mermin,rabson,physa,volc}, and even extended to treat
novel long-range order possessing color~\cite{rmpcolor} or magnetic
symmetry~\cite{prlmag,led}); and many fundamental problems---including
Bak's famous question: ``Where are the atoms?''~\cite{bakatoms}---are
gradually finding their solutions.  Nevertheless, other questions that
are equally important have remained unanswered to this day. Some of
these---such as the stabilization of quasicrystals, the role of
clusters, and the importance of the phason degree of freedom---were
hotly debated at a recent International Conference on
Quasicrystals~\cite{icq9}.

We know today that quasicrystals are more common than one had
originally expected. Scores of binary and ternary {\it metallic
  alloys\/} are known to form quasicrystalline
phases~\cite{tsai}---mostly with icosahedral or decagonal point-group
symmetry---and more are being discovered all the time.  Nevertheless,
it is only in the last couple of years that quasicrystals have been
discovered (independently) in two different {\it soft-matter\/}
systems: dendrimers~\cite{zeng1,zeng2,mehl} and star block
copolymers~\cite{takano1,matsushita,takano2}.  These newly discovered
{\it soft quasicrystals\/} not only provide exciting alternative
experimental platforms for the basic study of quasiperiodic long-range
order, but also hold the promise for new applications based on
self-assembled nanomaterials~\cite{scaffolds,electronics,%
  selfassemblyreview}, with unique electronic or photonic properties
that take advantage of the quasiperiodicity. One example is the
isotropic photonic band gaps that have been demonstrated in
artificially-constructed octagonal~\cite{jin} and
dodecagonal~\cite{zoorob} quasicrystals. 

The first step in a theoretical study of soft quasicrystals would be
to provide an explanation for their thermodynamic stability. To this
date, soft quasicrystals have been observed only with dodecagonal
point-group symmetry. Their source of stability is therefore likely to
be different from their solid-state siblings, yet a good understanding
of the stability of one quasiperiodic system may help to understand
the stability of the other. The purpose of this article is to propose
initial suggestions as to the source of stability of soft
quasicrystals, while providing a concise background on the subject.

\section{Quasicrystals---Terminology and general framework}

Let us consider a scalar function $\rho(\rv)$ that describes the
electronic density or the ionic potential of a material.  The Fourier
transform of a quasiperiodic density $\rho(\rv)$ (we shall assume here
that the transform always exists) has the form
\begin{equation}
  \label{eq:fourier}
  \rho(\rv) = \sum_{\kv\in L} \rho(\kv) e^{i\kv\cdot\rv},
\end{equation}
where the (reciprocal) lattice $L$ is a finitely generated
$\Zset$-module, {\it i.e.}  it can be expressed as the set of all
integral linear combinations of a finite number $D$ of $d$-dimensional
wave vectors, $\bv^{(1)},\ldots,\bv^{(D)}$. In the special case where
$D$, called the rank of the crystal, is equal to the physical
dimension $d$, the crystal is periodic. We refer to all quasiperiodic
crystals that are not periodic as ``quasicrystals''.\footnote[1]{Some
  older texts require crystals to possess so-called ``forbidden
  symmetries'' in order to be regarded as quasicrystals. It is now
  understood that such a requirement is inappropriate.  See
  Ref.~\cite{quasidef} for details, and Ref.~\cite{fibonacci} for
  examples of square and cubic quasicrystals.}

It is useful to introduce a physical setting based on the notion of
symmetry breaking~\cite{anderson,martin,sethna}. Let us assume that
the quasiperiodically-ordered state, described by $\rho(\rv)$, is a
symmetry-broken stable ground state of some generic free energy $\cal
F$, invariant under all translations and rotations in $\Rset^d$.
This is the same as saying that the physical interactions giving rise
to the quasicrystal are themselves translationally and rotationally
invariant, and that the ground state breaks this symmetry.
The free energy $\cal F$ is a functional of $\rho(\rv)$, which in
Fourier space takes the general form
\begin{equation}
  \label{eq:free}
  {\cal F}\{\rho\} = \sum_{n=2}^\infty \sum_{\kv_1\ldots\kv_n}
  A(\kv_1,\ldots\kv_n) \rho(\kv_1)\cdots\rho(\kv_n).
\end{equation}
Based on the idea of such a generic free energy, Rokhsar, Wright, and
Mermin~\cite{mermin} introduced the notion of {\it
  indistinguishability,} namely that two functions $\rho(\rv)$ and
$\rho'(\rv)$ are indistinguishable if a generic free energy cannot
distinguish between them and assigns them both the same value.  It
then follows that $\rho(\rv)$ and $\rho'(\rv)$ are indistinguishable
if and only if
\begin{equation}
  \label{eq:gauge}
  \forall \kv\in L:\quad \rho'(\kv) = e^{2\pi i\chi(\kv)}\rho(\kv),
\end{equation}
where $\chi(\kv)$, called a {\it gauge function,} has the property
that $\chi(\kv_1+\kv_2) \equiv \chi(\kv_1) + \chi(\kv_2)$ whenever
$\kv_1$ and $\kv_2$ are in $L$, where `$\equiv$' denotes equality to
within an additive integer.

Gauge functions are useful in describing the relations between the
different symmetry-broken ground states of $\cal F$. Dr\"ager and
Mermin~\cite{jorg} showed that gauge functions form a vector space
$V^*$ of all real-valued linear functions on the lattice $L$, and
because $L$ has rank $D$, $V^*$ is a $D$-dimensional vector space over
the real numbers. The space $V^*$ contains, as a subspace, all the
integral-valued linear functions on $L$. This subset, which has the
algebraic structure of a rank-$D$ $\Zset$-module (just like $L$
itself) is denoted by $L^*$. Gauge functions in $L^*$ leave the
ground-state density invariant. Gauge functions that belong to the
quotient space $V^*/L^*$ take the ground state described by $\rho$
into a different, yet indistinguishable, ground state described by
some other density function $\rho'$. Thus, one can parameterize all
the related symmetry-broken ground states of $\cal F$ on a simple
$D$-torus---the order parameter space $V^*/L^*$.

Different, yet indistinguishable, ground states may also be related by
rotations $g\in O(d)$. In this case $\rho'$ in (\ref{eq:gauge}) is
simply a rotated version of $\rho$, and for each such rotation $g$
there is a special gauge function $\phi_g$, called a {\it phase
  function,} satisfying
\begin{equation}
  \label{eq:point}
  \forall \kv\in L:\quad \rho(g\kv) = e^{2\pi i\phi_g(\kv)}\rho(\kv).
\end{equation}
The set of all rotations satisfying (\ref{eq:point}) forms the point
group of the crystal, and along with the corresponding phase functions
completely characterizes its space
group~\cite{rabson,mermin,physa,rmpcolor,prlmag,led}.

\section{Ordered soft matter}
\label{soft}

A wide variety of materials in physical-chemical and biological
systems lie in the intermediate region between simple liquids and
crystalline solids. Examples include liquid crystals, suspensions,
emulsions, polymer solutions or melts, surfactant solutions, and
biomaterials \cite{Witten}. These materials, termed {\it complex
  fluids}, {\it structured fluids}, or {\it soft condensed matter},
are mostly viscoelastic liquids, which contain a certain degree of
inner structure. Unlike simple fluids or solids, where either entropy
(in the former) or interactions (in the latter) dominate, the behavior
of soft matter is determined by an intricate interplay between
interactions and thermal fluctuations. Consequently, soft-matter
systems exhibit remarkably rich structures and dynamics, which have
been the subject of extensive experimental and theoretical research
during the past several decades
\cite{Witten,Wennerstrom,BenShaul,Safran,Gompper}.

Soft materials may possess long-range or quasi-long-range order
\cite{Chaikin,deGennes}. As was first demonstrated by Onsager
\cite{Onsager}, such liquid-crystalline phases can be obtained in
systems as simple as molecular liquids whose molecules have
anisotropic shapes. In more complex, self-assembling systems,
liquid-crystalline phases are usually made of intermediate structures,
which form as a result of incompatible molecular groups linked
together by chemical bonds. For example, amphiphilic molecules
(surfactants) are composed of hydrophilic and hydrophobic groups,
leading, in aqueous solution, to the formation of mesoscopic
structures such as micelles or fluid membranes
\cite{Wennerstrom,BenShaul,Safran}. These structures, at sufficiently
high concentration, organize into various ordered phases
\cite{Safran,Gompper}. Another important example is block
copolymers---macromolecules composed of two or more chemically bonded,
chain-like blocks \cite{Hamley1,Hamley2,Bates1,Bates2,Bates3}.  The
incompatibility of the different blocks gives rise to a variety of
ordered structures in melts and in solutions. The main advantage of
block copolymers as self-assembling systems is the relative ease at
which the sizes and properties of the different blocks (and, thus, the
resulting mesoscopic structures) can be controlled. Block copolymers,
like homopolymers, can be linear or branched. Two branched
architectures which can be well controlled, and thus have been drawing
considerable attention recently, are star polymers
\cite{Grest,asym_star,Lowen1,Lowen2,Lowen3,Lowen4,Lowen5,Lowen6,
  Lowen7,Lowen8,Kamien3}, where the chains are joined in one point,
and dendrimers \cite{Lowen7,Rubinstein1,Rubinstein2,percec2,Lowen9},
having a tree-like structure.

Phenomenological models based on generic free energies similar to
(\ref{eq:free}) have been very successful in describing phase diagrams
and transitions in soft matter
\cite{Chaikin,deGennes,Gompper,Bates2,Bates3,haim1,haim2}.  In such
models one identifies an order parameter, which may be a scalar
function---as in Eq.~(\ref{eq:free})---or a vector function, and
formulates, based on symmetry, a free-energy functional as an
expansion in the order parameter and its gradients. Such an expansion
is generally expected to become a good approximation close to a
continuous phase transition, where variations in the order parameter
occur over large length scales. The reason for the success of this
approach for soft systems, in particular, lies in their intermediate,
mesoscopic building blocks, which are significantly larger than the
atomic scale.  For example, spatial variations of the order-parameter
in block copolymer phases must occur over lengths larger than a
polymer block, which may be many nanometers. As a result, composition
variations and interfaces are smoother in soft materials than in
crystalline solids, rendering a long-wavelength gradient expansion a
much better approximation.  The situation becomes even better in the
case of flexible star polymers and dendrimers---which are of particular
relevance here---as the effective pair potentials in these
systems have an ``ultra-soft'' short-range repulsion
\cite{Lowen1,Lowen2,Lowen3,Lowen4,Lowen7,Lowen9,Kamien3}.

\section{Recent discovery of soft-matter quasicrystals}

Our current discussion is motivated by the recent experimental
discovery that soft matter can self-assemble into structures with
quasiperiodic long-range order.\footnote[2]{For the sake of historical
  accuracy, it should be noted that at some point the blue phase III
  of liquid crystals, also known as the ``blue fog'', was thought to
  have icosahedral quasicrystalline order~\cite{hornreich,rokhsar},
  but this eventually turned out not to be the
  case~\cite{wright,lubensky1}. Also, incommensurate helical
  twist-grain-boundary phases are known to exist in smectic liquid
  crystals~\cite{renn,goodby}, but the quasiperiodic order in this
  case is essentially only along the 1-dimensional screw axis.} In one
case, dendrimers that assume a conical shape assemble into micelles,
which then pack to form a perfect dodecagonal (12-fold)
quasicrystal~\cite{zeng1,zeng2,mehl}. In another case, $ABC$
star-shaped block terpolymers---in which the length ratios of the
three arms, $B/A$ and $C/A$, can be chemically-controlled---assemble
into a host of 2-dimensional columnar structures. One of these,
composed of square and triangle motifs (tiles), is suspected as being
a dodecagonal quasicrystal, or at least a periodic approximant of such
a quasicrystal, which under minor parameter tweaking may indeed form a
perfect quasicrystal~\cite{takano1,matsushita,takano2}. This phase has
also been reproduced numerically using lattice Monte Carlo simulations
by Dotera and Gemma~\cite{doteraICQ9}. A similar square-triangle
tiling has also been observed in a liquid crystal composed of T-shaped
molecules~\cite{chen}, which forms yet a third soft system which may
potentially self-assemble into a dodecagonal quasicrystal. The
characteristic length of the basic building blocks ranges in these
systems from $\sim10$ to $\sim100$ nanometers---2 to 3 orders of
magnitude greater than the atomic length scales found in hard
quasicrystals. This property of soft quasicrystals is what will
potentially make them useful as functional self-assembled
nanomaterials, and at the same time as a new experimental platform for
detailed---real-space and real-time---study of quasiperiodic
long-range order.

Very little is known at this point about these soft quasicrystals. For
example, even the space groups of the observed phases have not been
determined, although from the diffraction patterns of the dendrimer
liquid crystals given by Zeng {\it et al.}~\cite{zeng1,zeng2} it seems
that they have a 12-fold screw axis, and therefore, most likely, the
nonsymmorphic space group $P12_6/mcm$~\cite{rabson}.  More generally,
the same questions~\cite{icq9} concerning the mechanism of
stabilization, the role of clusters in formation and dynamics, and the
importance of phasons, apply to soft quasicrystals as they do to hard
quasicrystals.  Yet the answers may be more tractable (albeit possibly
different as the systems are quite different). Thus, the study of soft
quasicrystals will clearly have implications well beyond the limits of
the specific soft systems that have been discovered so far, and is
likely to promote the fundamental understanding of quasicrystals in
general.

\section{Why are soft quasicrystals stable? -- Insight from the
  Lifshitz-Petrich equation} 
\label{LP}

Motivated by experiments with parametrically-excited surface waves
(Faraday waves), exhibiting dodecagonal quasiperiodic
order~\cite{edwards}, Lifshitz and Petrich~\cite{faraday} developed a
model for describing the pattern-forming dynamics of a two-dimensional
field in which two length scales undergo a simultaneous instability.
This model is an extension of the Swift-Hohenberg
equation~\cite{swift}, which is used for describing a variety of
different pattern-forming systems~\cite{cross}.  Its dynamics is
relaxational, $\partial_t \rho = -\delta {\cal F} / \delta \rho$,
driving a 2-dimensional field $\rho(x,y,t)$ towards the minimum of an
``effective free energy'' (\ref{eq:free}),
\begin{equation}
 \label{eq:lyapunov}
 {\cal F}_{LP}\{\rho\} = \int \!dx\, dy\, \bigl\{- \frac12 \varepsilon
 \rho^2 + \frac12 
 [(\nabla^2+1)(\nabla^2+q^2)\rho]^2 
 - \frac13 \alpha \rho^3 + \frac14 \rho^4 \bigr\},
\end{equation}
yielding a dynamical equation of the form
\begin{equation}
 \label{lpeqn}
 \partial_t \rho = \varepsilon \rho - (\nabla^2 + 1)^2(\nabla^2 +q^2)^2 \rho
 +\alpha \rho^2 - \rho^3.
\end{equation}
It essentially mimics the dynamics of a generic 2-dimensional material
in search of its ground state, and therefore offers us important
insight and a good starting point for our current investigation of
soft quasicrystals.

The Lifshitz-Petrich free energy ${\cal F}_{LP}$ is indeed generic,
imposing only two requirements on a material, described by a
2-dimensional density $\rho(x,y,t)$: (a) The existence of two
characteristic length scales, whose ratio is given by the parameter
$q$; and (b) The existence of effective 3-body interactions, whose
importance is given by the relative strength of the parameter
$\alpha$. In~\cite{faraday} we were able to show analytically (using
standard methods~\cite[ch.~4.7]{Chaikin} and~\cite{gronlund}), and
demonstrate numerically, that if $q$ is chosen around $2\cos(\pi/12) =
\sqrt{2+\sqrt3} \simeq 1.932$ one can obtain a ground state with
quasiperiodic long-range order and dodecagonal
symmetry,\footnote[3]{Note the seeming coincidence that the
  star-shaped polymers, giving rise to the approximant dodecagonal
  quasicrystal, have ratios $B/A=1$ and
  $C/A=1.9$~\cite{matsushita,takano2}.} yet no choice of $q$ yields
globally-stable ground states with octagonal or decagonal symmetry.
The latter two have insufficient triplets of wave vectors in the
Fourier Lattice $L$ [Eq.~\(eq:fourier)] that add up to zero to
overcome the cost of additional density modes, as compared with the
hexagonal state. Thus, in two dimensions, the requirements of two
length scales and 3-body interactions are sufficient to stabilize
dodecagonal quasicrystals, but insufficient to stabilize octagonal or
decagonal quasicrystals.  This raises the possibility that the fact
that the soft quasicrystals discovered to date are all dodecagonal,
may be accounted for using a free energy similar to ${\cal F}_{LP}$.
Note that for hard quasicrystals the situation is
different---decagonal quasicrystals are thermodynamically stable
whereas octagonal and dodecagonal quasicrystals are believed to be
metastable---indicating that the stabilization mechanism for soft
quasicrystals might be quite different from that of hard
quasicrystals.

\section{Future directions}

\subsection{Free-energy functionals and stability analysis}
\label{sec.free}

At the outset, the experimental soft systems in which quasicrystalline
order has been observed seem to satisfy the basic assumptions of the
Lifshitz-Petrich theory described in Sec.~\ref{LP}. The asymmetric and
heterogeneous structure of the star polymers and dendrimers will most
likely require more than one length scale for an appropriate
coarse-grained description. Their ultra-soft repulsion and resulting
strong inter-penetration
\cite{Lowen1,Lowen2,Lowen3,Lowen4,Lowen7,Lowen9,Kamien3} imply that
3-body interactions should be significant \cite{Lowen6}.  Thus, we
expect that studies that we are currently undertaking will yield
functionals similar in nature to ${\cal F}_{\rm LP}$ of
Eq.~(\ref{eq:lyapunov}).  Significant differences may emerge,
nonetheless, as the systems considered here are 3-dimensional and
differ in their microscopic structure.  For instance, two order
parameters rather than one might be required \cite{takano2}, which
could potentially allow point-group symmetries other than dodecagonal
to be observed.\footnote[4]{Models with two order parameters were
  suggested also for hard quasicrystals~\cite{Mermin85} and
  pattern-forming systems~\cite{Muller}, yielding additional
  ground-state symmetries.}

A 3-dimensional version of an LP-like free energy may remind the
reader of the early attempts by Kalugin, Kitaev, and
Levitov~\cite[KKL]{kalugin}, who extended the model of Alexander and
McTague~\cite{alexander}, to establish that the icosahedral
quasicrystal has lower free energy than the competing {\it bcc\/}
phase.  Narasimhan and Ho~\cite[NH]{narasimhan} managed to show in
their model that there are regions in parameter space in which a
dodecagonal quasicrystal is favored and other regions in which a
decagonal quasicrystal is favored. These attempts were eventually
discontinued after it was shown by Gronlund and Mermin~\cite{gronlund}
that the addition of a quartic term to the cubic free energy of KKL
reverses the outcome of the calculation, establishing the {\it bcc\/}
phase as the favored one. For hard crystals it is unclear where to
truncate the expansion of the free energy and whether such a
truncation is fully justified. As has been mentioned in
Sec.~\ref{soft}, for the soft systems considered here the truncation
of the expansion should be more valid.  We therefore intend to
reexamine some of the old conclusions when considering the stability
of \emph{soft} quasicrystals based on generic free energy
considerations.  Roan and Shakhnovich~\cite{roan} performed such a
study for the case of icosahedral order in diblock copolymers and
concluded that such order is only metastable.  Nevertheless, we are
encouraged by the old results of NH who established the stability of
dodecagonal, as well as decagonal, quasicrystals within the same
model.

Another key insight can be drawn from a recent theoretical
observation, according to which dispersions of soft, fuzzy, particles
are essentially different in their thermodynamics from those of hard
particles \cite{Kamien1,Kamien2}. The overlap of the soft ``coronas''
surrounding the particles leads to a driving force acting to minimize
their interfacial area, in analogy with foams.  Consequently, unusual
liquid-crystalline structures can be stabilized in systems of soft
spheres \cite{Kamien1,Kamien2,li,Kamien4}. Both star polymers and
flexible dendrimers fall into this fuzzy category, yet they may be
highly aspherical. Likos {\it et al.}~\cite{Lowen9} have also shown
that stars and flexible dendrimers have the same kind of soft pair
potentials.  We thus expect such considerations of interfacial-area
minimization to become highly relevant in the upcoming study of soft
quasicrystals.

\subsection{Dislocation and phason dynamics}
\label{sec.dislocations}

Valuable knowledge about the nature of quasiperiodic order can be
obtained by studying its topological
defects~\cite{anderson,sethna,mermindefects}, and its low-energy
collective excitations---in particular those associated with the
phason degrees of freedom. Much like phonons, phasons are low-energy
excitations of the quasicrystal, only that instead of slightly
shifting the atoms away from their equilibrium positions, the relative
positions of atoms are changed. Their existence stems directly from
the fact that the dimension $D$ of the order parameter space $V^*/L^*$
is greater than the physical dimension $d$. Thus, in addition to $d$
independent (acoustic) phonon modes there are $D-d$ independent phason
modes. The exact role of phasons in the stabilization and the dynamics
of quasicrystals is not fully-understood~\cite{icq9}, yet their
presence is unequivocally-detected in numerous types of experiments.
Edagawa {\it et al.}~\cite{edagawa} have recently managed to observe
the real-time thermal fluctuations of single phason-flips in a
solid-state quasicrystal---a remarkable achievement that requires a
very sophisticated experiment.  Soft quasicrystals, with their larger
length scales, may be much better suited for real-time investigation
of phason dynamics.

We have recently begun investigating the motion of dislocations and
the dynamics of phasons in the dodecagonal ground state of the LP
equation~\cite{barak1,barak2}. We are studying, both analytically and
numerically, such questions as the climb velocity of dislocations
under strain, the pinning of dislocations by the underlying
quasiperiodic structure under conditions of weak diffusion, and the
relaxation of phason strain as two dislocations of opposite
topological sign merge and annihilate each other. We have also used
similar theoretical procedures in the analysis of defect dynamics in a
real experimental nonlinear photonic quasicrystal~\cite{nature}.
Similar investigations of more realistic models of soft quasicrystals,
or on the soft systems themselves, should provide valuable
insight into their physical nature.

\section*{Acknowledgments}

One of the authors (RL) thanks X. Zeng and T. Dotera for fruitful
discussions while attending ``Aperiodic 2006'' in Zao, Japan. This
research is funded by the Israel Science Foundation through grant
number 684/06.

\begin{singlespacing}

\end{singlespacing}

\end{document}